\documentclass[twocolumn,english,amsmath,amssymb,superscriptaddress,showpacs]{revtex4}
\usepackage[T1]{fontenc}
\usepackage[latin9]{inputenc}
\usepackage{amsmath}
\usepackage{graphicx}
\usepackage{amssymb}
\usepackage{esint}

\makeatletter
\@ifundefined{textcolor}{}
{%
 \definecolor{BLACK}{gray}{0}
 \definecolor{WHITE}{gray}{1}
 \definecolor{RED}{rgb}{1,0,0}
 \definecolor{GREEN}{rgb}{0,1,0}
 \definecolor{BLUE}{rgb}{0,0,1}
 \definecolor{CYAN}{cmyk}{1,0,0,0}
 \definecolor{MAGENTA}{cmyk}{0,1,0,0}
 \definecolor{YELLOW}{cmyk}{0,0,1,0}
 }

\makeatother

\makeatother

\makeatother

\makeatother

\makeatother

\makeatother

\makeatother

\makeatother

\usepackage{babel}

\makeatother

\usepackage{babel}

\begin{document}

\title{Electron-Phonon Coupling in High-Temperature Cuprate Superconductors
Determined from Electron Relaxation Rates}

\author{C. Gadermaier}

\email[To whom correspondence should be addressed: ]{christoph.gadermaier@ijs.si}

\affiliation{Department of Complex Matter, Jozef Stefan Institute, Jamova 39,
1000 Ljubljana, Slovenia}

\author{A. S. Alexandrov}

\affiliation{Department of Physics, Loughborough University, Loughborough LE11
3TU, United Kingdom}

\affiliation{Department of Complex Matter, Jozef Stefan Institute, Jamova 39,
1000 Ljubljana, Slovenia}

\author{V. V. Kabanov}

\affiliation{Department of Complex Matter, Jozef Stefan Institute, Jamova 39,
1000 Ljubljana, Slovenia}

\author{P. Kusar}

\affiliation{Department of Complex Matter, Jozef Stefan Institute, Jamova 39,
1000 Ljubljana, Slovenia}

\author{T. Mertelj}

\affiliation{Department of Complex Matter, Jozef Stefan Institute, Jamova 39,
1000 Ljubljana, Slovenia}

\author{X. Yao}

\affiliation{Department of Physics, Shanghai Jiao Tong University, Shanghai 200240,
China}

\author{C. Manzoni}

\affiliation{National Laboratory for Ultrafast and Ultraintense Optical Science,
INFM-CNR, Dipartimento di Fisica, Politecnico di Milano, 20133 Milano,
Italy}

\author{D. Brida}

\affiliation{National Laboratory for Ultrafast and Ultraintense Optical Science,
INFM-CNR, Dipartimento di Fisica, Politecnico di Milano, 20133 Milano,
Italy}

\author{G. Cerullo}

\affiliation{National Laboratory for Ultrafast and Ultraintense Optical Science,
INFM-CNR, Dipartimento di Fisica, Politecnico di Milano, 20133 Milano,
Italy}

\author{D. Mihailovic}

\affiliation{Department of Complex Matter, Jozef Stefan Institute, Jamova 39,
1000 Ljubljana, Slovenia}
\begin{abstract}
We determined electronic relaxation times via pump-probe optical spectroscopy
using sub-15 fs pulses for the normal state of two different cuprate
superconductors. We show that the primary relaxation process is the
electron-phonon interaction and extract a measure of its strength,
the second moment of the Eliashberg function $\lambda\langle\omega^{2}\rangle=800\pm200$
meV$^{2}$ for La$_{1.85}$Sr$_{0.15}$CuO$_{4}$ and $\lambda\langle\omega^{2}\rangle=400\pm100$
meV$^{2}$ for YBa$_{2}$Cu$_{3}$O$_{6.5}$. These values suggest
a possible fundamental role of the electron-phonon interaction in
the superconducting pairing mechanism. 
\end{abstract}


\maketitle
The electron-phonon interaction (EPI) is decisive for determining
the functional properties of materials. It is the main scattering
process governing electronic conductivity and is crucial for the formation
of ordered electronic states such as charge-density waves and often
the superconducting state. The determination of its strength - usually
defined as the second moment $\lambda\langle\omega^{2}\rangle=2\int_{0}^{\infty}\alpha^{2}F(\omega)\omega d\omega$
of the Eliashberg spectral function $\alpha^{2}F(\omega)$ \cite{Eliashberg}
- is thus of fundamental importance. Standard methods for determining
$\lambda\langle\omega^{2}\rangle$ experimentally from phonon linewidths
in Raman or neutron scattering are often biased by selection rules
and inhomogeneous broadening, and have given controversial results
in the past. Since scattering from phonons is one of the main relaxation
processes for electrons, $\lambda\langle\omega^{2}\rangle$ can be
accurately extracted from the electron-phonon relaxation time $\tau_{e-ph}$,
provided that: (i) the experiment affords adequate time resolution
to determine $\tau_{e-ph}$, and (ii) an appropriate model connecting
$\lambda\langle\omega^{2}\rangle$ and $\tau_{e-ph}$ is used. We
will show in this paper that for materials with strong EPI, to satisfy
both conditions, we need to go beyond current approaches. Here, by
using optical spectroscopy with ultra-high time-resolution (< 20 fs
instrument response) and a new, more appropriate model, we obtain
$\lambda\langle\omega^{2}\rangle$ values for two high-critical temperature
($T_{c}$) cuprate superconductors, which allows us to assess the
role of the EPI in the superconducting mechanism in these materials.
Ultrahigh time-resolution is important to detect fast processes in
strongly interacting systems, and to correctly identify the EPI relaxation
in cases where the data contain the dynamics of several processes.
Since for strong EPI $\tau_{e-ph}$ can be well below 100 fs, we need
a better resolution than the usual 50-80 fs used so far \cite{elbert,chek,perfetti,brorson}.
Therefore we use state-of-the art ultrashort laser pulses from two
synchronised non-collinear optical parametric amplifiers \cite{manzoni}.

In femtosecond optical pump-probe spectroscopy the sample is excited
with a short pump laser pulse, and the reflectivity is measured with
a (weaker) probe pulse at a variable delay. The pump beam is periodically
modulated and the photoinduced signal is expressed as a relative change
of the reflected light intensity $\frac{\Delta R}{R}=\frac{R_{pump}-R_{0}}{R_{0}}$
, where $R_{pump}$ and $R_{0}$ are the reflected intensities with
and without pump pulse, respectively. The temporal evolution of $\frac{\Delta R}{R}$,
which - for small perturbations - is related to the temporal evolution
of the dielectric constant $\Delta\epsilon/\epsilon$, is a direct
signature of the energy relaxation processes in the sample. We used
15-fs pump pulses centred at 530 nm and broad band sub-10-fs probe
pulses with a spectrum ranging from 500 to 700 nm (a detailed scheme
is found in the supplementary information). This non-degenerate pump-probe
configuration eliminates coherent interference artefacts. Single crystals
of YBa$_{2}$Cu$_{3}$O$_{6.5}$ (YBCO, $T_{c}$ = 60 K) and La$_{1.85}$Sr$_{0.15}$CuO$_{4}$
(LSCO, $T_{c}$ = 38 K) were prepared as in ref. \cite{yao,primoz}.
To avoid any competing relaxation processes from emergent low temperature
states (e.g. superconducting, pseudogap, antiferromagnetic, or stripe
order), we performed all experiments at room temperature.

Until recently, a theoretical framework expressing $\tau_{e-ph}$
in terms of $\lambda\langle\omega^{2}\rangle$ has been provided by
the so-called two-temperature model (TTM) \cite{kag,allen}. It is
based on the assumption that the relaxation time due to electron-electron
(e-e) collisions $\tau_{e-e}$ is much shorter than $\tau_{e-ph}$.
The e-e scattering is assumed to establish a thermal distribution
of electrons with a temperature $T_{e}$ > $T_{l}$ ($T_{l}$ being
the lattice temperature) on a time scale typically faster than the
experimental time resolution. The relaxation time $\tau_{e-ph}$ of
subsequent electron cooling via EPI is related to $\lambda\langle\omega^{2}\rangle$:

\begin{equation}
\lambda\langle\omega^{2}\rangle=\frac{\pi}{{3}}\frac{k_{B}T_{e}}{{\hbar\tau_{e-ph}}}\label{exact}\end{equation}

This expression has been used in the analysis of transient optical
experiments \cite{elbert,chek,brorson}, and recently also time-resolved
angle-resolved photoemission spectroscopy (ARPES) \cite{perfetti}.
For the typical laser fluences used in these experiments $T_{e}$
is in the range 400-800 K . This gives an estimate for $\tau_{e-e}=$
350 fs $\div$ 1.4 ps, depending on the fluence (see supplementary
information). From Eq. (1), $\tau_{e-ph}$ is expected to be proportional
to $T_{e}$ and thus vary significantly over the range of fluences
used in the experiment.

Outside the TTM regime, the relaxation behavior can be described via
the kinetic Boltzmann equation using e-e and e-ph collision integrals,
where the electrons and phonons are both out of equilibrium. This
has been done both numerically \cite{lag} and recently also analytically
\cite{gusev,alekab}. The calculated electron distribution based on
the analytical solution of this non-equilibrium model (NEM) \cite{alekab}
departs from the equilibrium Fermi-function particularly for high
energies. (A comparison with published time-resolved ARPES data is
shown in the supplement.) The analytical treatment yields a relation:

\begin{equation}
\lambda\langle\omega^{2}\rangle=\frac{2\pi}{{3}}\frac{k_{B}T_{l}}{{\hbar\tau_{e-ph}}}.\label{exact}\end{equation}

which is applicable also when $\tau_{e-e}$ > $\tau_{e-ph}$. Besides
the factor 2, a notable difference compared to the TTM formula (Eq.
(1)) is that $\tau_{e-ph}$ is predicted to be linearly dependent
on $T_{l}$ (not $T_{e}$). Since the heat capacity of the lattice
is much higher than that of electrons, in our experiments $T_{l}$
is close to room temperature for all fluences, so we expect that $\tau_{e-ph}$
should be independent of fluence. This provides a critical test of
the model's applicability.

The experimental data for LSCO ( Figs 1a+b)) and YBCO (Figure 1c))
show a fast initial decay followed by a slower dynamics, all of which
are independent of laser fluence. We fit the transient response of
both cases (see Figs 1b+c) with two exponential decays with time constants
$\tau_{a}$ and $\tau_{b}$ respectively, and a long-lived plateau,
using the pump-probe cross-correlation as the generation term. For
each sample, the same $\tau_{a}$ and $\tau_{b}$ are obtained at
different probe wavelengths. YBCO also contains an oscillatory response
due to impulsively excited coherent phonons. This coherent phonon
contribution can be removed almost entirely by fitting the oscillatory
response of the known Raman-active modes and subtracting it from the
data (see Fig. 1c)). The fact that in YBCO at 520 nm the two signal
components have opposite sign nicely confirms that we are actually
observing two processes and not at a non-exponential process which
could accidentally be fitted with two exponentials. The fit yields
$\tau_{a}=45\pm8$ fs and $\tau_{b}=600\pm100$ fs for LSCO and $\tau_{a}=100\pm20$
fs and $\tau_{b}=450\pm100$ fs for YBCO respectively. This behavior
is systematically observed over the whole spectral range of our probe
pulse between 500 and 700 nm (see supplementary information).

\begin{figure}
\includegraphics[bb=100bp 110bp 385bp 680bp,clip,width=7cm]{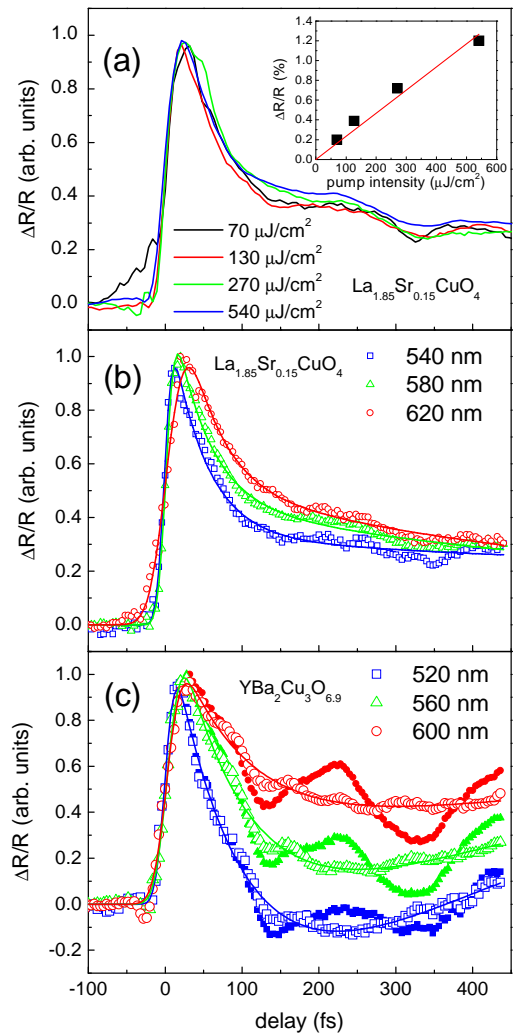}

\caption{a) Normalised photoinduced reflectivity change in La$_{1.85}$Sr$_{0.15}$CuO$_{4}$
at 590 nm for different pump intensities. The inset shows the signal
magnitude as a function of pump intensity. b) $\Delta R/R$ of La$_{1.85}$Sr$_{0.15}$CuO$_{4}$
at different probe wavelengths (symbols) and double-exponential fits
(lines). c) $\Delta R/R$ of YBa$_{2}$Cu$_{3}$O$_{6.5}$ at different
probe wavelengths and double-exponential fits. Small full symbols
show the original data, large open symbols show the data after subtraction
of three oscillating modes at 115, 145, and 169 cm$^{-1}$. These
oscillations arise from a modulation of the reflectivity by phonons
coherently excited in the sample by the pump pulse, whose duration
is much shorter than the oscillation period \cite{chwalek,kutt}.
The three modes are known from Raman spectroscopy \cite{heyen,iliev}.}

\end{figure}

Since the observed dynamics is fluence independent (see Figure 1a),
neither of the two fast relaxation processes can be attributed to
$e-e$ scattering. Following previous studies \cite{perfetti,elbert},
we assign $\tau_{a}$ to relaxation via the EPI mechanism. The origin
of the longer relaxation time $\tau_{b}$ has been discussed in detail
previously \cite{primoz,perfetti,elbert,Kabanov} and is of no further
interest here.

The choice of model (TTM or NEM) for determining $\lambda\langle\omega^{2}\rangle$
from $\tau_{e-ph}$ is based on the relatively stringent requirement
regarding the fluence dependence of $\tau_{e-ph}$. A fluence-dependent
$\tau_{e-ph}$ is clearly not observed here, and to the best of our
knowledge has never been observed in cuprates. We thus conclude that
the TTM is not applicable, while the data are consistent with the
NEM solution without the assumption that $\tau_{e-e}<<\tau_{e-ph}$.
Calculating the EPI strength, Eq. (2) yields $\lambda\langle\omega^{2}\rangle=800\pm200$
meV$^{2}$ for LSCO and $\lambda\langle\omega^{2}\rangle=400\pm100$
meV$^{2}$ for YBCO. As additional confirmation regarding the choice
of model, we note that a dependence of $\tau_{e-ph}$ on the sample
temperature, as predicted by the NEM \emph{has} actually been observed
in cuprates \cite{toda} and superconducting iron pnictides \cite{pnic}
above the pseudogap temperature, where it is expected to apply. No
such dependence is predicted by the TTM.

To assess the consequences of using the NEM rather than the traditional
TTM, in the supplementary information we compare $\lambda\langle\omega^{2}\rangle$
values obtained with the two models both for our data and for several
metals from the literature. The TTM assumption $\tau_{e-e}<<\tau_{e-ph}$
is generally not valid. The discrepancy in $\lambda\langle\omega^{2}\rangle$
calculated with the two models can be up to a factor of 2. If different
fluences are used, as in our data, the variation of $T_{e}$ introduces
an additional uncertainty if we use the TTM estimate.

To obtain an estimate of $\lambda$ from the data, we express the
second moment of the Eliashberg function as the product of a dimensionless
electron-phonon coupling constant $\lambda$ and the square of a characteristic
phonon frequency $\omega_{0}$: $\lambda\langle\omega^{2}\rangle=\lambda\omega_{0}^{2}$.
The estimate of $\omega_{0}$ and consequently $\lambda$ requires
a detailed knowledge of the Eliashberg spectral function. This can
be extracted from other experiments such as optical absorption \cite{mic,zamb},
neutron scattering \cite{arai,ega,rez}, ARPES \cite{lanzara,meev},
and tunnelling \cite{ved,zas,zhou,lee,shim,zhao}. Based on these
references the best estimate of $\omega_{0}$ is about 40 meV, which
gives $\lambda\gtrsim0.5$ for LSCO and $\lambda\gtrsim0.25$ for
YBCO. Remarkably, these values agree very well with ab initio calculations
that predict 0.27 for YBCO \cite{Bohnen} and 0.4 for LSCO \cite{giustino,sangiovanni}.

To assess the possible contribution of EPI to the superconductive
pairing mechanism in the cuprates, we briefly discuss the observations
in terms of existing theories based on phonon mediated pairing \textendash{}
most notably BCS theory and polaronic pairing \cite{ale83,ale88,MK,tomaz}.
BCS theory predicts that $k_{B}T_{c}=\hbar\omega_{0}\exp[-(1+\lambda)/\lambda]$
(if any repulsive Coulomb pseudopotential is neglected). At maximum
($\lambda$=2, $\omega_{0}$=$\sqrt{\lambda\langle\omega^{2}\rangle/2}$,
see supplementary information) the BCS critical temperature can be
$T_{c}^{max}=$ 52 K for LSCO and only $T_{c}^{max}=$ 37 K for YBCO.
Remarkably - contrary to the experiment - it predicts a \emph{higher}
$T_{c}$ for LSCO than for YBCO.

Polaronic pairing within the band picture, on the other hand, yields
a maximum $T_{c}(\lambda)$ that is significantly higher than for
BCS and is obtained at a lower $\lambda$ value. Polaronic band-narrowing
due to phonon \textquotedbl{}dressing\textquotedbl{} of carriers strongly
enhances the density of states in a narrow polaron band and consequently
also the critical temperature of polaronic superconductors \cite{ale83}.
With further increase of the EPI strength carriers form real-space
(bipolaronic) pairs and the critical temperature, which is now the
Bose-Einstein condensation temperature, drops since the effective
mass of these composed bosons increases \cite{ale88}. The highest
$T_{c}(\lambda)$ exceeding the BCS value by several times is hence
found in the intermediate crossover region of the EPI strength from
the weak-coupling BCS to the strong-coupling polaronic superconductivity.
Strong e-e correlations increase the effective mass of carriers (or
decrease the bare band-width), and heavier carriers form lattice polarons
at a smaller value of $\lambda$ \cite{feh,mis} ($\lambda_{c}\thickapprox0.9$
for uncorrelated 2D polarons \cite{kab}, while $\lambda_{c}\lesssim0.4$
in the Holstein t-J model \cite{mis}). The observed EPI strengths
are therefore consistent with polaronic pairing in the presence of
strong electron correlations, whereby YBCO lies in the crossover region
close to the maximum $T_{c}$, while LSCO would appear to be on the
strong-coupling side of this region ($\lambda>\lambda_{c}$). Alternatively,
within local bipolaron pairing models \cite{MK}, the limits of $T_{c}$
are set by (dynamic or static) phase coherence percolation \cite{MKM},
where the interplay of the EPI and the Coulomb repulsion between doped
carriers $V_{c}$ determine the pair density and detailed real-space
texture \cite{tomaz}. These models give a charge-ordered regime when
Coulomb repulsion dominates ($\lambda/N_{0}\ll V_{c}$, with $N_{0}$
being the density of states at the Fermi energy) and a fully phase
separated state when EPI is dominant ($\lambda/N_{0}\gg V_{c}$).
In the crossover region between these two regimes, a textured state
favoring pair (bipolaron) formation exists, leading to superconductivity
with a distinct maximum $T_{c}$.

Our results reinforce the other compelling experimental evidence for
a strong role for the EPI in cuprates obtained from isotope effects
\cite{zhao}, high resolution ARPES \cite{lanzara,meev}, optical
\cite{mic,zamb}, neutron-scattering \cite{arai,ega,rez}, and tunnelling
\cite{lee,shim,zhaoPRL} spectroscopies. However, our data on two
materials can only demonstrate the realistic feasibility of the polaronic
pairing mechanism, and cannot rule out any non-phononic contribution
to the pairing. Indeed part of the glue function has been identified
with an energy well above the upper limit of the phonon frequencies
in the cuprates (100 meV) \cite{marel}. While this could be a signature
of multi-phonon dressing of carriers, spin and/or electron density
fluctuations might be alternative mechanisms of the high-energy glue.
By using the appropriate theory and adequate time resolution, as we
have shown, one can now collect accurate data for further cuprate
high-$T_{c}$ materials to decide whether the agreement with the polaronic
mechanism is coincidental or systematic. Similar work will be of fundamental
significance for other effects where EPI is important, such as high
$T_{c}$ superconductivity in non-cuprate materials (notably iron-pnictides
\cite{pnic}), colossal magnetoresistance, the formation of orbitally-ordered
states and charge density waves. 
\begin{acknowledgments}
This work was supported by the Slovenian Research Agency (ARRS) (grants
J1-2305, 430-66/2007-17 and BI-CN/07-09-003), EPSRC (UK) and the Royal
Society (grants EP/D035589/1 and JP090316), MOST of China (Project
2006CB601003), and by the European Commission {[}grants EIF-040958A,
ERG-230975, and the European Community Access to Research Infrastructure
Action, Contract RII3-CT-2003-506350 (Centre for Ultrafast Science
and Biomedical Optics, LASERLAB-EUROPE){]}. We thank S. Sugai for
the La$_{1.85}$Sr$_{0.15}$CuO$_{4}$ sample and D. Polli and L.
Stojchevska for fruitful discussions. \end{acknowledgments}

\end{document}


\title{Electron-phonon coupling in cuprate high-temperature superconductors
determined from exact electron relaxation rates. Supplementary material.}

\author{C. Gadermaier$^{1}$, A. S. Alexandrov$^{2,1}$, V. V. Kabanov$^{1}$,
P. Kusar$^{1}$, T. Mertelj$^{1}$, X. Yao$^{3}$, C. Manzoni$^{4}$,
D. Brida$^{4}$, G. Cerullo$^{4}$, and D. Mihailovic$^{1}$ }

\affiliation{$^{1}$Josef Stefan Institute 1001, Ljubljana, Slovenia\\
 $^{2}$Department of Physics, Loughborough University, Loughborough
LE11 3TU, United Kingdom\\
 $^{3}$Department of Physics, Shanghai Jiao Tong University,
800 Dong Chuan Road, Shanghai 200240, China\\
 $^{4}$National Laboratory for Ultrafast and Ultraintense Optical
Science, INFM-CNR, Dipartimento di Fisica, Politecnico di Milano,
20133 Milano, Italy }

\maketitle

\subsection{Sample preparation}

The YBa$_{2}$Cu$_{3}$O$_{6.5}$ (YBCO) single crystal used here
was grown by top-seeded solution growth using a Ba$_{3}$Cu$_{5}$O
solvent\cite{Yao}. The as-grown single crystal was first annealed
at 700 \textdegree{}C for 70 h with flowing oxygen and quenched down
to room temperature. The La$_{1.85}$Sr$_{0.15}$CuO$_{4}$ (LSCO)
single crystal was synthesised by a travelling-solvent-floating-zone
method utilizing infrared radiation furnaces (Crystal system, FZ-T-4000)
and annealed in oxygen gas under ambient pressure at 600 \textdegree{}C
for 7 days\cite{sugai}.

\subsection{Femtosecond pump-probe set-up}

A detailed description of the set-up used is found in\cite{manzoni}.
In a \textquotedblleft{}pump-probe\textquotedblright{} experiment,
a \textquotedblleft{}pump\textquotedblright{} pulse excites the sample
and the induced change in transmission or reflection of a delayed
probe pulse monitors the relaxation behaviour. In the linear approximation
$\frac{\Delta R}{R}$ directly tracks the electronic relaxation processes,
and the time constants obtained from fits of its dynamics are the
characteristic times of the underlying relaxation processes. In our
data, this approximation is justified by two essential characteristics:
(i) the $\frac{\Delta R}{R}$ amplitude is linear in the excitation
intensity (see Figure 1a for LSCO), and (ii) the same decay times
appear independently of the probe wavelength, only with different
spectral weights of the individual components.

In order to resolve the dynamics of fast processes very short pulses
are necessary, since the instrumental response function is given by
the cross correlation between the pump and probe pulses. We use sub-10
fs probe pulses from an ultrabroadband (covering a spectral range
from 500 to 700 nm) non-collinear optical parametric amplifier (NOPA)
and $\thicksim$15 fs pump pulses from a narrower band (wavelength
tunable, in our case centred at 530 nm) NOPA. The seed pulses for
the NOPAs and the amplified pulses are steered and focussed exclusively
with reflecting optics to avoid pulse chirping.

\begin{figure}
\includegraphics[bb=85bp 485bp 525bp 725bp,width=70mm]{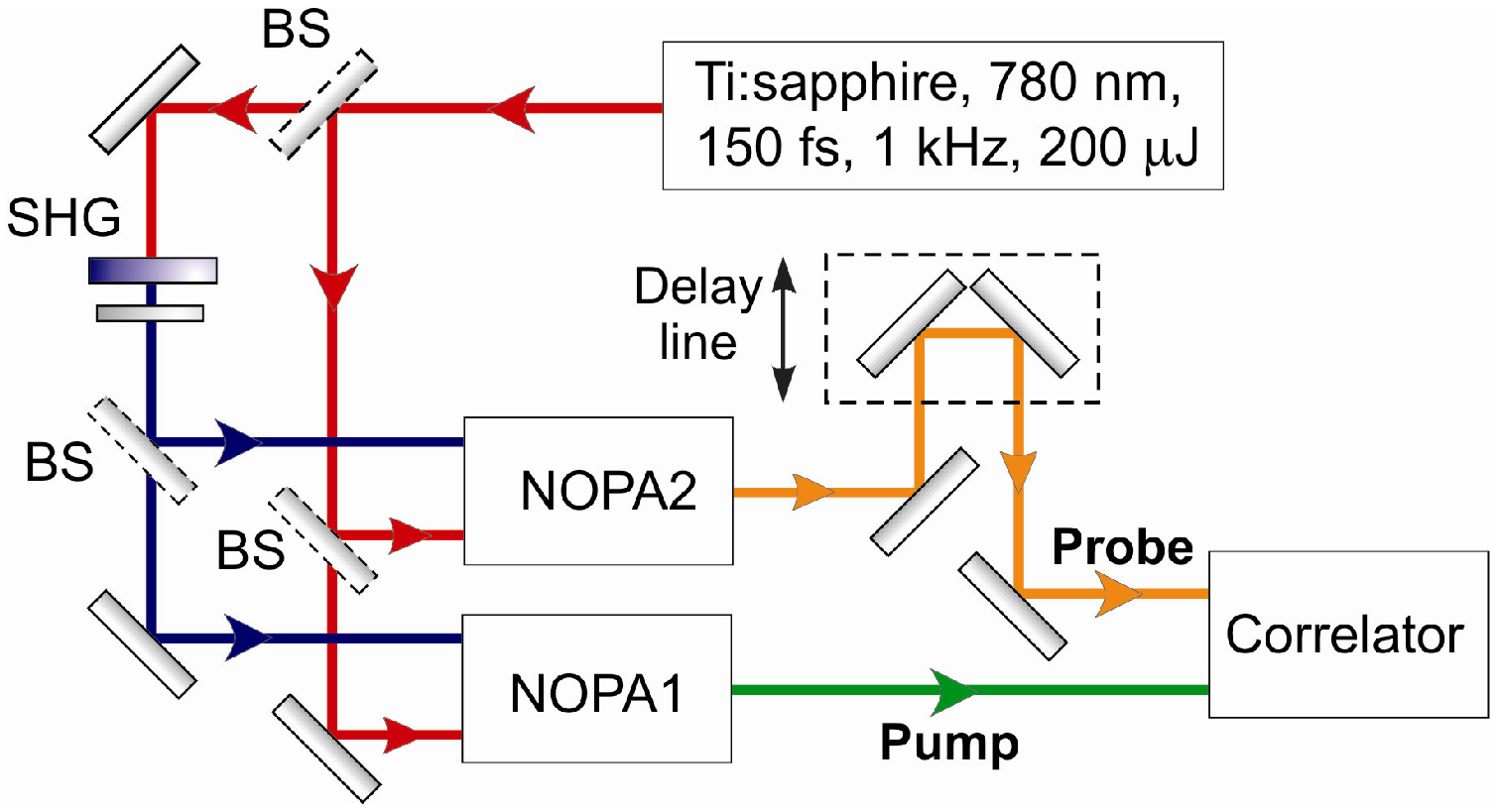}\caption{Block scheme of the experimental setup. BS: beam splitter. SHG: second
harmonics generation. (redrawn from \cite{manzoni}).}

\end{figure}

A schematic of the experimental apparatus is shown in Fig. 1. The
laser source is a regeneratively amplified modelocked Ti:sapphire
laser (Clark-MXR Model CPA-1), delivering pulses at 1 kHz repetition
rate with 780 nm center wavelength, 150 fs duration, and 500 $\mu$J
energy. Both NOPAs are pumped by the second harmonic of the Ti:sapphire
laser, which is generated in a 1-mm-thick lithium triborate crystal
(LBO), cut for type-I phase matching in the XY plane ($\theta$ =
90\textdegree{}, $\varphi$ = 31.68\textdegree{}, Shandong Newphotons).

\begin{figure}
\includegraphics[bb=85bp 515bp 525bp 725bp,width=70mm]{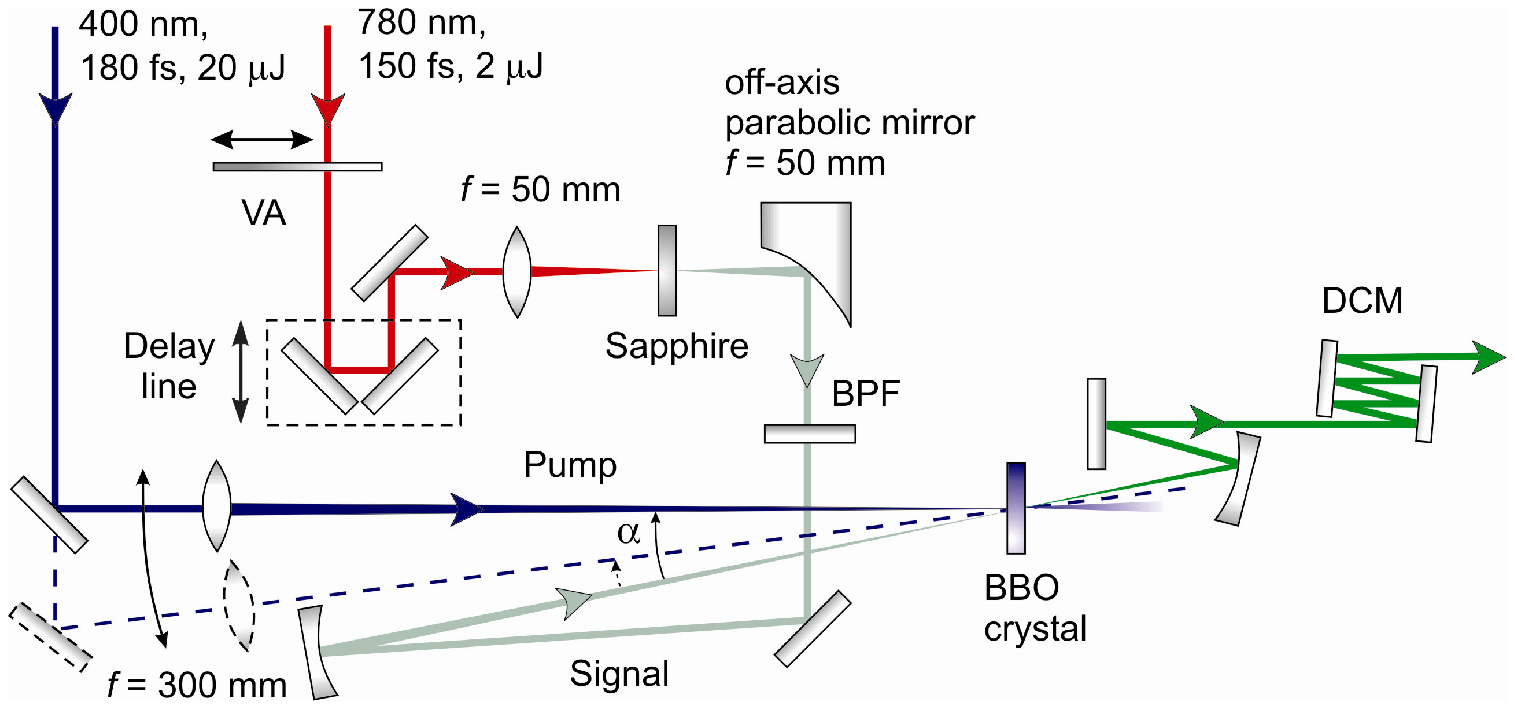}\caption{Setup of the noncollinear optical parametric amplifier. BPF: short
pass filter that transmits only the visible part of the white light
continuum and cuts out the fundamental and infrared components. (redrawn
from \cite{manzoni}).}

\end{figure}

The ultrabroadband visible NOPA that generates the probe pulses has
been described in detail before\cite{giulio}; a schematic of it
is shown in Fig. 2. The white light continuum seed pulses are generated
by a small fraction of the fundamental wavelength beam focused into
a 1 mm thick sapphire plate. Parametric gain is achieved in a 1-mm-thick
BBO crystal, cut at $\theta$= 32\textdegree{}, which is the angle
giving the broadest phase matching bandwidth for the noncollinear
type-I interaction geometry; a single-pass configuration is used to
maximize the gain bandwidth. The amplified pulses have energy of approximately
2 $\mu$J and peak-to-peak fluctuations of less than 5\%.

The compressor for the ultrabroadband NOPA consists of two custom-designed
double-chirped mirrors (DCMs), manufactured by ion-beam sputtering
(Nanolayers GmbH), which are composed of 30 pairs of alternating SiO2
/TiO2 quarter-wave layers in which both the Bragg wavelength and the
layer duty cycle are varied from layer pair to layer pair. The DCMs
introduce a highly controlled negative group delay (GD) over bandwidths
approaching 200 THz, compensating for the GD of the NOPA pulses.

The narrower bandwidth NOPA providing the pump pulses is built identically
to the one described above, only the amplified bandwidth is reduced
by choosing a suitable non-optimum angle between pump and signal beams
(dashed path of the blue pump in Figure 2). The pulses are compressed
by DCMs similar to those used for the broadband NOPA.

\begin{figure}
\includegraphics[bb=85bp 310bp 525bp 725bp,width=70mm]{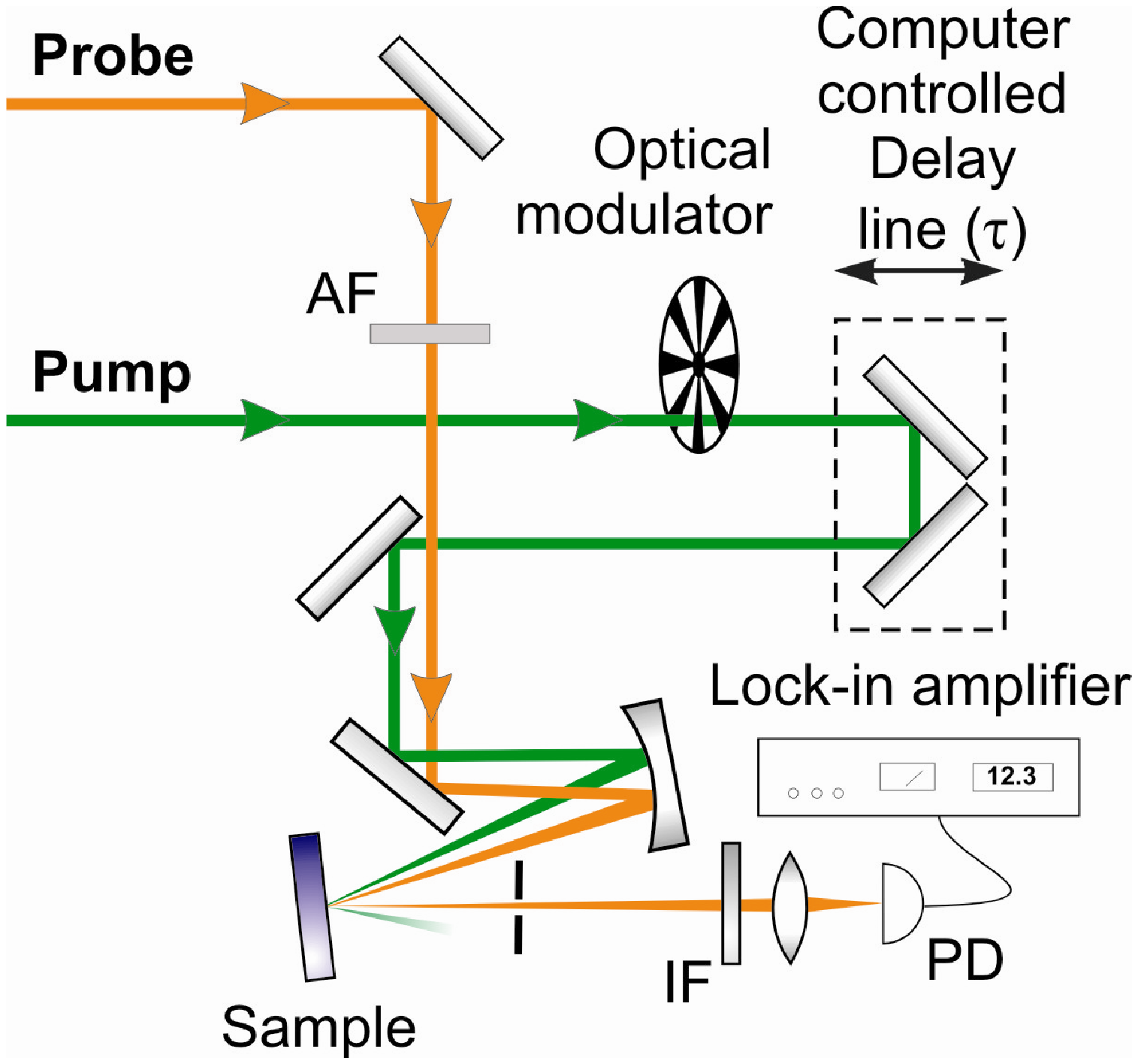}\caption{Schematics of the experimental apparatus used for auto/ cross-correlation
and pump-probe experiments. AF: attenuating filter. IF: interference
filter. PD: photodiode. (redrawn from \cite{manzoni}).}

\end{figure}

A schematic of the apparatus used for pulse characterization and pump-probe
experiments (correlator in Fig. 1) is shown in Fig. 3. The delay line
is formed by two 90\textdegree{} turning mirrors mounted on a precision
translation stage with 0.1 $\mu$m positioning accuracy (Physik Instrumente
GmbH, model M-511.DD), which corresponds to 0.66 fs time resolution.
The two pulses are combined and focused on the sample by a silver
spherical mirror (R=200 mm). The non-collinear configuration enables
to spatially separate pump and probe beams. Upon reflection from the
sample, the probe beam is selected by an iris and steered to the detector,
either a silicon photodiode preceded by a 10 nm spectral width interference
filter or an optical multichannel analyzer (OMA). The differential
reflection (transmission) signal is obtained via synchronous detection
(lock-in amplifier Stanford SR830 for the photodiode, custom made
software for the OMA) referenced to the modulation of the pump beam
at 500 Hz by a mechanical chopper (Thorlabs MC1000). This allows detection
of differential reflection ($\Delta$R/R) signals as low as 10$^{-4}$.

\subsection{Estimate of the electron-electron relaxation time}

The applicability of the FL theory has been somewhat controversal
in cuprate superconductors. The recent unambiguous observation of
de Haas \textendash{} van Alphen oscillations\cite{hussey} has shown
that the Fermi surface is almost cylindrical. In this quasi-two-dimensional
case the electrons can be described as a FL if $r_{s}=a/a_{B}<37$,
with $a$ being the mean electron distance and $a_{B}$ the Bohr radius\cite{tanatar}.
We determine $a=1/\sqrt{n}=\sqrt{2\pi}/k_{F}$, with $k_{F}=$ 7.4
nm$^{-1}$ from\cite{hussey}. Using $a_{B}=\hbar^{2}\epsilon_{eff}/m^{*}e^{2}$,
with $m^{*}=4m_{e}$ and the effective dielectric constant $\epsilon_{eff}\approx30$,
we obtain $r_{s}\approx1$, safely in the FL regime.

Compared to a simple FL, e-e correlations increase the effective mass
of carriers (or decrease the bare band-width), and heavier carriers
form lattice polarons at a smaller value of l. Both spin and lattice
polarons have the same Fermi surface as free electrons, but the Fermi
energy is reduced \cite{para}. Our (Boltzmann) relaxation theory
is based on the existence of the Fermi surface and the Pauli exclusion
principle, which the dressed (polaronic) carriers obey like free electrons.
Therefore, Equation 3 of the main paper and all our subsequent considerations
are valid also for polarons in the presence of strong electron correlations.

In the weak photoexcitation regime where only a small fraction of
conduction electrons is excited ($k_{B}T_{e}\ll E_{F}$), e-e scattering
is impeded since the Pauli exclusion principle strongly limits the
number of available final states. The e-e relaxation time $\tau_{e-e}$
can be estimated by: $1/\tau_{e-e}\approx\pi^{3}\mu_{c}^{2}(k_{B}T_{e})^{2}/4\hbar E_{F}$
\cite{Ashcroft}, where $\mu_{c}=r_{s}/2\pi$ is the Coulomb pseudopotential
characterising the electron-electron interaction. Even in the weak
photoexcitation regime, the effective $T_{e}$ after excitation can
differ significantly from room temperature. Assuming an electronic
specific heat of 1.4 mJ/g.at.K$^{2}$ \cite{loram}, and a pump laser
penetration depth of 150 nm, one can estimate $T_{e}\thickapprox$
400 K, $\tau_{e-e}\thickapprox$ 1.4 ps for the lowest and $T_{e}\thickapprox$
800 K, $\tau_{e-e}\thickapprox$ 350 fs for the highest pump fluence
used in Fig. 1a of the main manuscript.

\subsection{Exact relaxation rates}

Here, differently from previous studies based on the two-temperature
model (TTM), we analyze pump-probe relaxation rates using an analytical
approach to the Boltzmann equation \cite{kabAlex}, which is free
of any quasi-equilibrium approximation. Due to the complex lattice
structure of cuprate superconductors characteristic phonon frequencies
spread over a wide interval $\hbar\omega/k_{B}\gtrsim200\div1000$
K. Very fast oxygen vibrations with frequencies $\omega\gtrsim0.1/fs$
do not contribute to the relaxation on the relevant time scale, but
just dress the carriers. For the remaining part of the spectrum we
can apply the Landau-Fokker-Planck expansion, expanding the e-ph collision
integral at room or higher temperatures in powers of the relative
electron energy change in a collision with a phonon, $\hbar\omega/(\pi k_{B}T)\lesssim1$.
Then the integral Boltzmann equation for the nonequilibrium part of
the electron distribution function $\phi(\xi,t)=f(\xi,t)-f_{0}(\xi)$
is reduced to a partial differential equation in time-energy space
\cite{kabAlex}: \begin{equation}
\gamma^{-1}\dot{\phi}(\xi,t)=\frac{\partial}{{\partial\xi}}\left[\tanh(\xi/2)\phi(\xi,t)+\frac{\partial}{{\partial\xi}}\phi(\xi,t)\right],\end{equation}
where $f(\xi,t)$ is the non-equilibrium distribution function, and
$\gamma=\pi\hbar\lambda\langle\omega^{2}\rangle/k_{B}T$. The electron
energy, $\xi$, relative to the equilibrium Fermi energy is measured
in units of $k_{B}T$. Here $\lambda\langle\omega^{2}\rangle$ is
the second moment of the familiar Eliashberg spectral function \cite{Eliashberg},
$\alpha^{2}F(\omega)$, defined for any phonon spectrum as: \begin{equation}
\lambda\langle\omega^{n}\rangle\equiv2\int_{0}^{\infty}d\omega\frac{\alpha^{2}F(\omega)\omega^{n}}{{\omega}}.\end{equation}
 The coupling constant $\lambda$, which determines the critical temperature
of the BCS superconductors, is $\lambda=2\int_{0}^{\infty}d\omega\alpha^{2}F(\omega)/\omega$.

Multiplying Eq.(1) by $\xi$ and integrating over all energies yield
the rate of the energy relaxation: \begin{equation}
\dot{E}_{e}(t)=-\gamma\int_{-\infty}^{\infty}d\xi\tanh(\xi/2)\phi(\xi,t),\end{equation}
 where $E_{e}(t)=\int_{-\infty}^{\infty}d\xi\xi\phi(\xi,t)$ and $\phi(\xi,t)$is
the solution of Eq. 1. Apart from a numerical coefficient the characteristic
e-ph relaxation rate (proportional to $\gamma$) is about the same
as the TTM energy relaxation rate \cite{allen}, $\gamma_{T}=3\hbar\lambda\langle\omega^{2}\rangle/\pi k_{B}T$.

\begin{figure}
\includegraphics[bb=105bp 485bp 380bp 690bp,width=70mm]{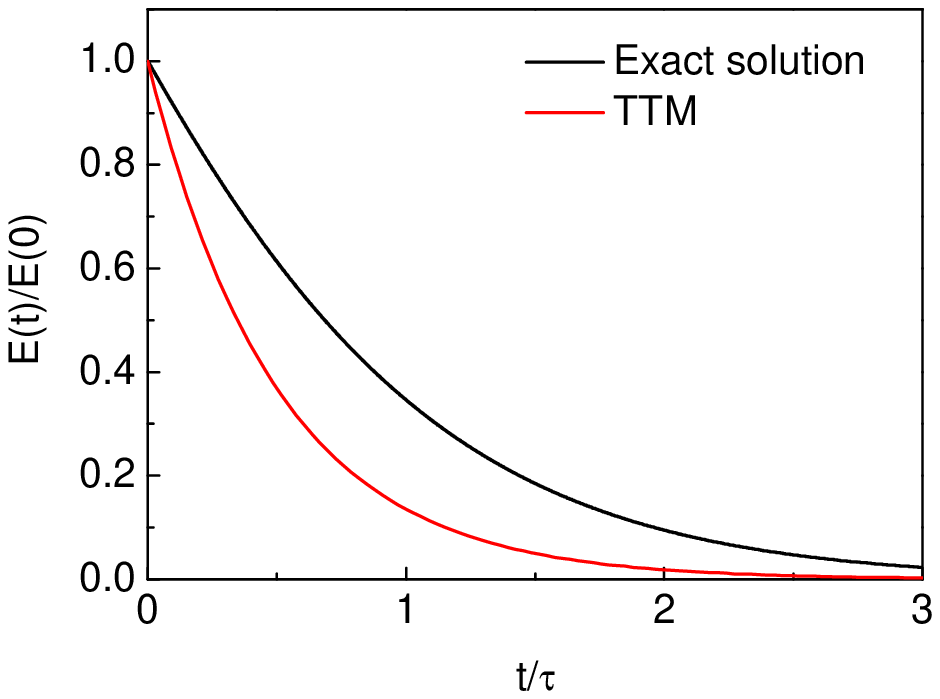}\caption{Numerically exact energy relaxation rate compared with TTM rate. $E_{0}$
is the pump energy, $\tau=2\pi k_{B}T/3\hbar\lambda\langle\omega^{2}\rangle$
is the exact relaxation time.}

\end{figure}

To establish the numerical coefficient we solved Eq.(1) and fitted
the numerically exact energy relaxation by a near-exponential decay
$E(t)=E_{0}\exp{(-at\gamma-bt^{2}\gamma^{2})}$, with the coefficients
$a=0.14435$ and $b=0.00267$, which are virtually independent of
the initial distribution, $\phi(\xi,0)$ and pump energy, $E_{0}$.
The exact relaxation time $\tau=1/a\gamma$ turns out longer than
the TTM relaxation time $\gamma_{T}^{-1}$, by a factor of 2, (see
Fig. 4), $\tau=2\pi k_{B}T/3\hbar\lambda\langle\omega^{2}\rangle$.
This as well as the shorter relaxation times observed in our ultra-fast
pump-probe measurements lead to essentially higher values of EPI coupling
constants compared with previous studies.

\subsection{Exact transient electron distribution function}

In the TTM e-e collision creates a quasi-equilibrium electron distribution
that subsequently cools down via e-ph interaction and electron diffusion.
On the other hand, in our relaxation scheme, which is dominated by
e-ph interaction, the electron distribution during relaxation is not
a quasi-equilibrium one. The \emph{deviation} of both functions from
the equilibrium is shown in Fig. 5, for $t=\gamma^{-1}$. Since we
limit our discussion to the linear regime, the TTM correction to the
distribution function is $\Delta f\varpropto x/cosh(x^{2}/4)$, where
$x=\xi/k_{B}T$. The width of the two lobes in $\Delta f$ indicate
the effective electronic temperature. The significantly narrower lobes
for the TTM compared to the exact distribution illustrate how the
TTM underestimates the transient electronic temperature and hence
the relaxation time. However, the e-ph dominated relaxation is not
just a slower relaxation through the same quasi-equilibrium states
as assumed by the TTM. We illustrate this by fitting our non-equilibrium
distribution with a quasi-equilibrium one (dashed line in Fig. 5).
Since the Fokker-Planck equation describes diffusion in the energy
space the high energy tails are clearly seen in our distribution,
which, compared to the Fermi-Dirac distribution have a much gentler
fall-off towards high electron energies.

\begin{figure}
\includegraphics[bb=105bp 480bp 375bp 690bp,width=70mm]{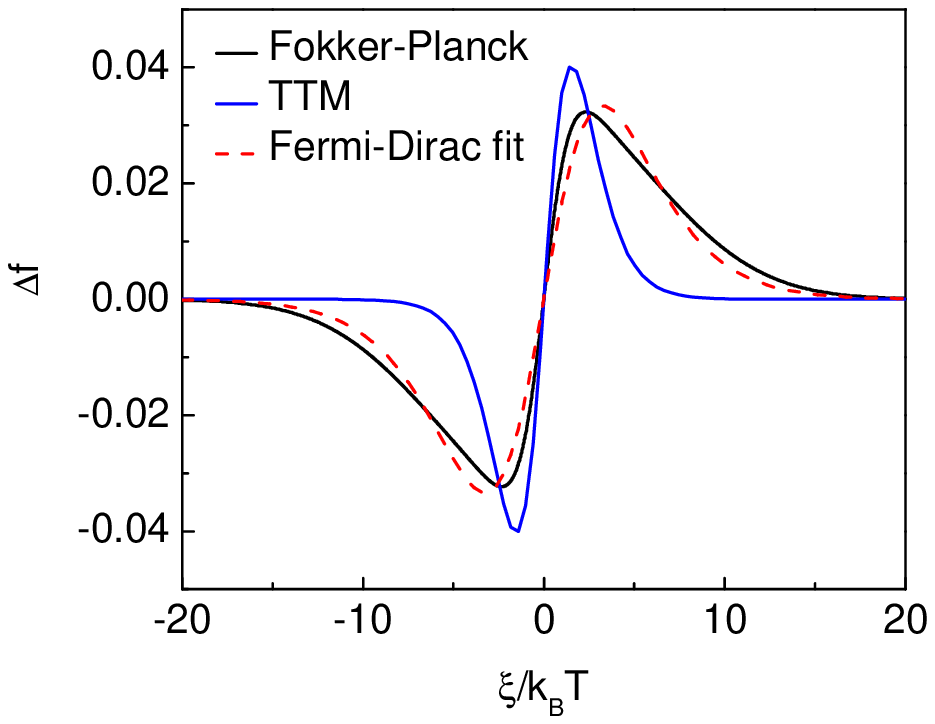}\caption{Difference $\Delta f=f(\xi,t=\gamma^{-1})-f_{0}(\xi)$ of the transient
distribution functions $f(\xi,t=\gamma^{-1})$ and the equilibrium
Fermi-Dirac function $f_{0}(\xi)$ . $f(\xi,t)$ is calculated from
the Fokker-Planck equation and via the TTM, respectively.}

\end{figure}

\begin{figure}
\includegraphics[bb=100bp 370bp 285bp 660bp,width=70mm]{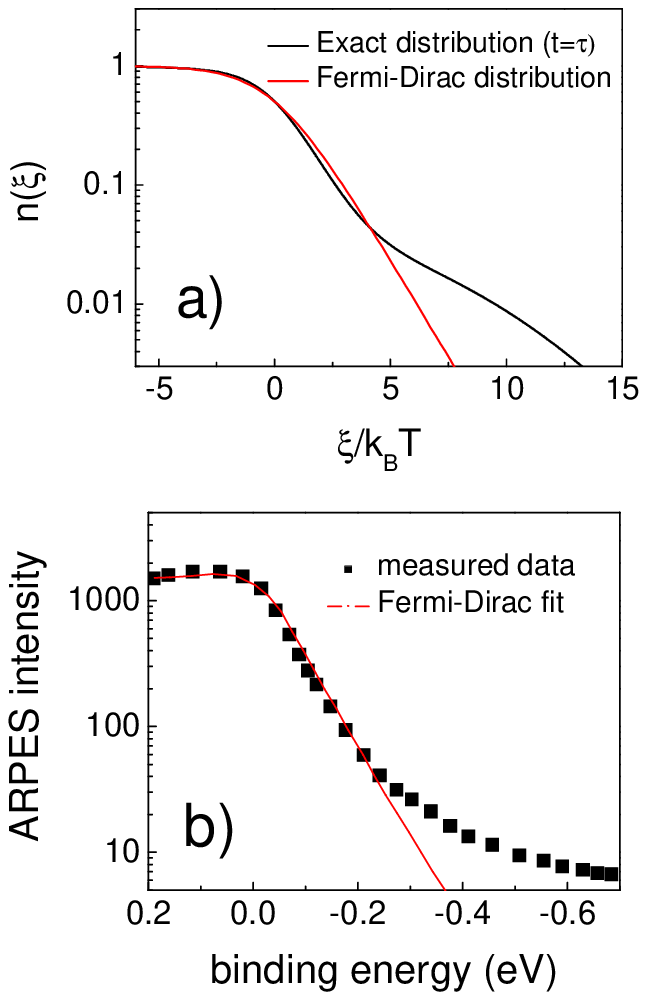}\caption{a) Numerically exact distribution function $f(\xi,t=\gamma^{-1})$
compared to a Fermi-Dirac distribution function with the same effective
temperature. b) ARPES spectrum of Bi$_{2}$Sr$_{2}$CaCu$_{2}$O$_{8}$immediately
after excitation and fit to a Fermi-Dirac function (redrawn from \cite{perfetti}).}

\end{figure}

While in the deviation from the equilibrium distribution, the difference
between the two models is clearly visible, the Fermi-Dirac distribution
and the one obtained from solving the Fokker-Planck equation, themselves
look rather similar, except for the high energy tail, as can be seen
in Fig. 6a. This shines new light on fs ARPES experiments, which directly
measure the transient electron distributions. Until now it was customary
to use the TTM to describe electron relaxation, ARPES data were in
good agreement and were invoked as confirmation of the ultrafast e-e
thermalization. We have now shown that a \emph{similar} distribution
can be obtained also as a result of e-ph scattering. In Fig. 6b we
redraw the best available time-resolved ARPES data for cuprate superconductors
\cite{perfetti} together with their fit to a quasi-equilibrium Fermi-Dirac
distribution, from which they estimate a hot electron temperature.
The spectrum was taken immediately after excitation (at the end of
the pump-probe pulse overlap, which is about $\tau/2$ of their decay
time after maximum pump-probe overlap, i.e. on average the electrons
are probed $\tau/2$ after their excitation). Compared to the Fermi-Dirac
curve, their data show a high-energy tail very similar to the exact
non-equilibrium distribution we calculated. This means that our relaxation
scheme is not only more justified than the TTM on the basis of the
physical reasoning described in the main text, it also agrees better
with experimental data in the literature. This should in no way derogate
the work done before our model was available, however, we propose
to reassess quantitative conclusions along the lines of our relaxation
scheme (see section G).

\subsection{Predictions of BCS theory}

\begin{figure}
\includegraphics[bb=75bp 55bp 580bp 675bp,angle=270,width=70mm]{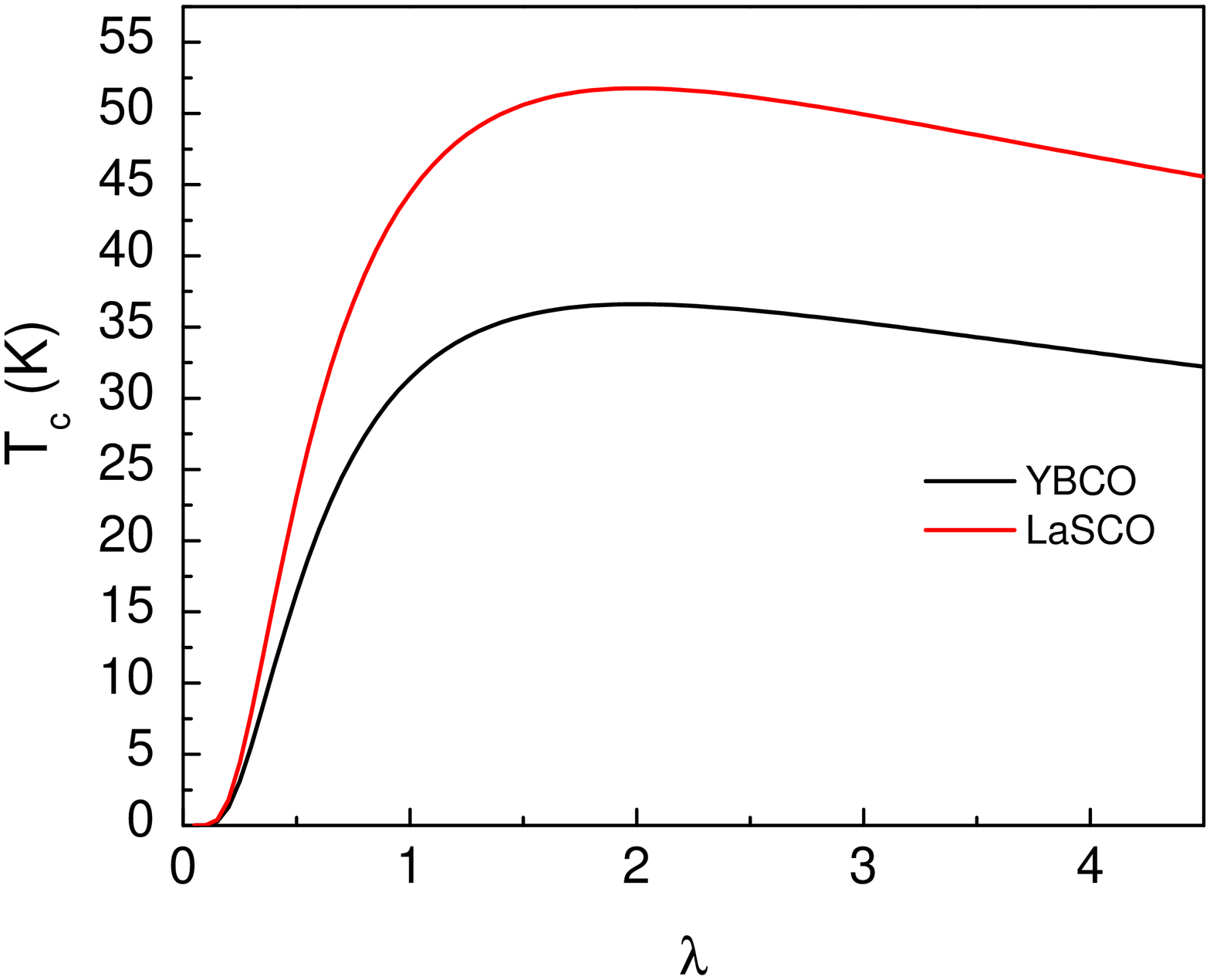}\caption{Critical temperature according to the BCS-McMillan formula as a function
of $\lambda$ for YBa$_{2}$Cu$_{3}$O$_{6.5}$ ($\lambda\langle\omega^{2}\rangle=400$
meV$^{2}$) and La$_{1.85}$Sr$_{0.15}$CuO$_{4}$ ($\lambda\langle\omega^{2}\rangle=800$
meV$^{2}$)}

\end{figure}

The most common expression that relates EPI to $T_{c}$ is the BCS-McMillan
formula $k_{B}T_{c}=\hbar\omega_{0}\exp[-(1+\lambda)/\lambda]$ (if
any repulsive Coulomb pseudopotential is neglected), where $\omega_{0}$
is a characteristic phonon frequency. Formally setting $\lambda\langle\omega^{2}\rangle=\lambda\omega_{0}^{2}$,
one can rewrite this as a function of $\lambda\langle\omega^{2}\rangle$
and $\lambda$: $k_{B}T_{c}=\hbar\sqrt{\lambda\langle\omega^{2}\rangle/\lambda}\exp[-(1+\lambda)/\lambda]$.
Since $\lambda\langle\omega^{2}\rangle$ is known from experiment,
we keep it fixed and find a maximum $T_{c}(\lambda)$ (by finding
a zero of the first derivative) at $\lambda=2$. $T_{c}$ as function
of $\lambda$ is shown in Fig 7 for both YBCO and LSCO, the maximum
critical temperatures are $T_{c}^{max}=$ 52 K for LSCO and $T_{c}^{max}=$
37 K forYBCO. For the more realistic estimates given in the main text,
$\lambda=0.5$ for LSCO and $\lambda=0.25$ for YBCO, we obtain lower
$T_{c}$ values of 23 K and 3 K, respectively. Hence, contrary to
experiment BCS theory predicts a lower $T_{c}$ for YBCO than for
LSCO. It cannot explain the high $T_{c}$ value of YBCO (even for
$\lambda=2$) and the reasonable agreement for LSCO is probably a
coincidence.

\subsection{Assessment of lambda values obtained with the TTM and with our model}

We now assess the quantitative differences between the values for
$\lambda\langle\omega^{2}\rangle$ (and consequently the estimates
for $\lambda$ that are usually made on their basis) obtained from
data analysis using the TTM and our model. The differences between
the analytic expressions that link $\tau_{e-ph}$ and $\lambda\langle\omega^{2}\rangle$
(equations 1 and 2 of the main manuscript) are a factor 2 and that
$\tau_{e-ph}$ scales with the electron temperature $T_{e}$ in the
TTM and with the lattice temperature $T_{l}$ in our model. As described
in section C, $T_{l}$ is usually very close to the sample temperature
without photoexcitation, while $T_{e}$ can be several 100 K higher,
depending on the excitation conditions. For the fluences we used,
$T_{e}$ = 600 $\pm$ 200 K, which introduces an additional uncertainty
in $\lambda\langle\omega^{2}\rangle$ if we use the TTM estimate.
The factor 2 in the equations incidentally cancels with the factor
2 in the temperatures.

\begin{table}
\caption{$\lambda\langle\omega^{2}\rangle$ obtained via the TTM and the Kabanov-Alexandrov
(K-A) model from our data on YBa$_{2}$Cu$_{3}$O$_{6.5}$ and La$_{1.85}$Sr$_{0.15}$CuO$_{4}$
and from \cite{brorson}. $\tau_{e-e}$ are calculated as in section
C, $\tau_{e-ph}$ from \cite{brorson} are recalculated from published
$\lambda\langle\omega^{2}\rangle$ and $T_{e}$ values. }

\begin{tabular}{llllll}
material  & $T_{e}$ (K) & $\tau_{e-e}$ (fs)  & $\tau_{e-ph}$ (fs)  & $\lambda\langle\omega^{2}\rangle_{TTM}$ (meV$^{2}$) & $\lambda\langle\omega^{2}\rangle_{K-A}$ (meV$^{2}$)\tabularnewline
\hline 
YBa$_{2}$Cu$_{3}$O$_{6.5}$ & 400-800 & 350-1400  & 100 & 400$\pm$150 & 400$\pm$100\tabularnewline
La$_{1.85}$Sr$_{0.15}$CuO$_{4}$ & 400-800  & 350-1400 & 45 & 800$\pm$300 & 800$\pm$200\tabularnewline
Cu & 590 & 3300 & 1400 & 29 & 29\tabularnewline
Au & 650 & 1700 & 1900 & 23 & 21\tabularnewline
Cr & 720 &  & 380 & 130 & 110\tabularnewline
W & 1200 &  & 710 & 110 & 60\tabularnewline
V & 700 &  & 170 & 280 & 240\tabularnewline
Nb & 790 & 1000 & 170 & 320 & 240\tabularnewline
Ti & 820 &  & 160 & 350 & 260\tabularnewline
Pb & 570 & 6400 & 840 & 45 & 47\tabularnewline
NbN & 1070  &  & 110 & 640 & 360\tabularnewline
V3Ga & 1110  &  & 200 & 370 & 200\tabularnewline
\hline
\end{tabular}
\end{table}

Brorson et al. studied a series of metallic superconductors, using
a quite diverse range of excitation conditions \cite{brorson}. We
list their data together with ours in Table 1. Their $T_{e}$ values
scatter from 570 to 1200 K, hence the differences between the TTM
and our model can be up to a factor of 2. Unfortunately they did not
study any intensity dependence, so we cannot use this criterion to
decide which relaxation scheme is more appropriate. However, we estimate
$\tau_{e-e}$ as described in section C for four of their samples,
using $r_{s}$ and $E_{F}$ from \cite{Ashcroft} and obtain $\tau_{e-e}>\tau_{e-ph}$
in three cases, and $\tau_{e-e}\approx\tau_{e-ph}$ for Au. Therefore,
one can assume that except for materials with very low EPI, our model
is more appropriate, as has already been established in textbooks
like \cite{Ashcroft}.

\subsection{Universality of the observed behavior}

In the main paper we show intensity and wavelength dependent pump-probe
time traces and show that the dynamics does not change with intensity
and that the same characteristic time scales are found at different
wavelengths. However, there we show only a few selected wavelengths
out of a more comprehensive dataset. To illustrate that the behavior
discussed in the main paper is universal, we show two-dimensional
maps of the photoinduced $\Delta R/R$ as a function of probe wavelength
and delay for both materials in Figure 8.

\begin{figure}
\includegraphics[bb=85bp 70bp 320bp 695bp,angle=270,width=11cm]{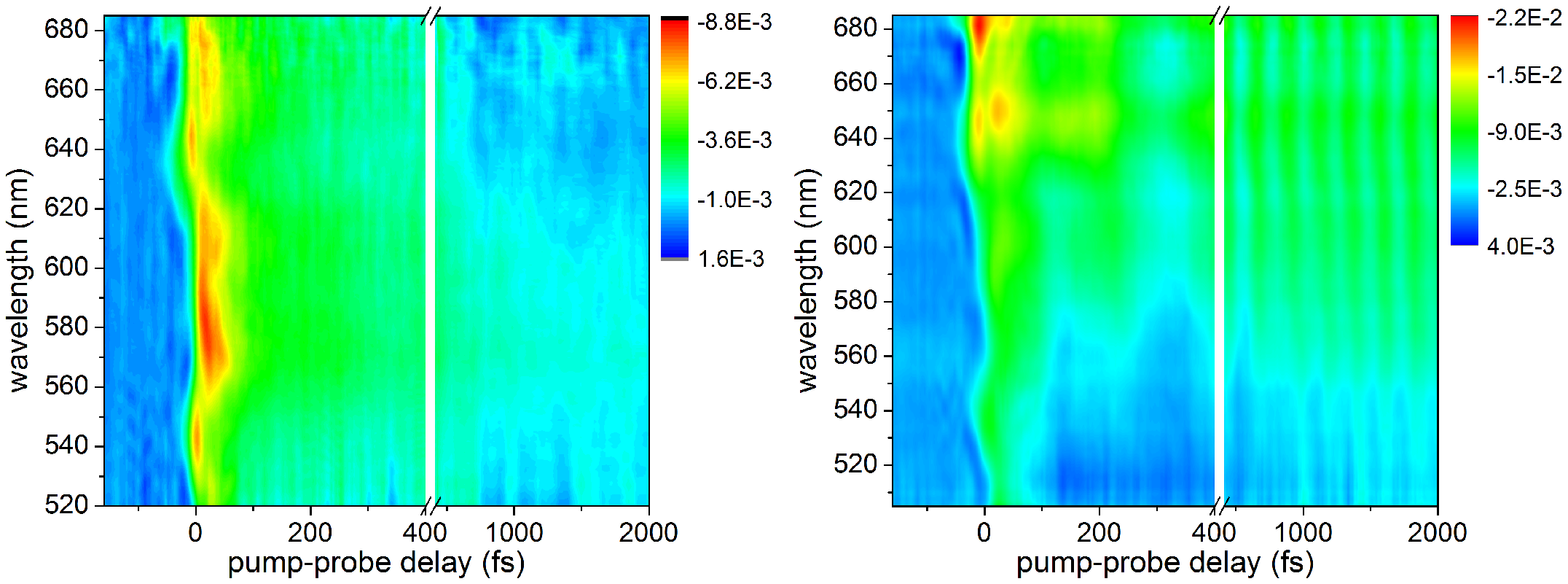}\caption{Transient photoinduced reflectivity change $\Delta R/R$ of La$_{1.85}$Sr$_{0.15}$CuO$_{4}$
(a) and YBa$_{2}$Cu$_{3}$O$_{6.5}$ (b) at a pump intensity of 200
$\mu$J/cm$^{2}$.}

\end{figure}